\documentclass[prd,twocolumn,floatfix,amsmath,nofootinbib,amssymb,floatfix]{revtex4}
\usepackage{graphicx,color,dcolumn,booktabs,bm}
\usepackage{longtable,lscape}
\usepackage{txfonts}
\usepackage{overpic}
\usepackage{indentfirst}
\usepackage{feynmf}   
\usepackage{slashed}  
\usepackage{cases}
\usepackage{float}
\usepackage{multirow}
\usepackage{ulem}
\usepackage{subfigure}
\usepackage{dsfont,amssymb,amsmath,amsfonts,amsbsy,mathrsfs}

\usepackage{hyperref}
\hypersetup{colorlinks,citecolor=blue,anchorcolor=red,menucolor=red, linkcolor=red,filecolor=red,runcolor=red,urlcolor=blue,frenchlinks=true}

\makeatletter
\@addtoreset{equation}{section}
\makeatother
\allowdisplaybreaks

\begin{document}

\title{Probing exotic $J^{PC}$ resonances from 
 deeply bound charmoniumlike  molecules:  Insights for identifying exotic hadrons}
\author{Zhi-Peng Wang$^{1,2}$}
\email{wangzhp2020@lzu.edu.cn}
\author{Fu-Lai Wang$^{1,2,3,4}$}
\email{wangfl2016@lzu.edu.cn}
\author{Guang-Juan Wang$^{5}$}
\email{wgj@post.kek.jp}
\author{Xiang Liu$^{1,2,3,4,6}$}
\email{xiangliu@lzu.edu.cn}
\affiliation{$^1$School of Physical Science and Technology, Lanzhou University, Lanzhou 730000, China\\
$^2$Research Center for Hadron and CSR Physics, Lanzhou University and Institute of Modern Physics of CAS, Lanzhou 730000, China\\
$^3$Key Laboratory of Quantum Theory and Applications of MoE, Lanzhou University,
Lanzhou 730000, China\\
$^4$Lanzhou Center for Theoretical Physics, Key Laboratory of Theoretical Physics of Gansu Province, Lanzhou University, Lanzhou 730000, China\\
$^5$KEK Theory Center, Institute of Particle and Nuclear Studies (IPNS), High Energy Accelerator Research
Organization (KEK), 1-1 Oho, Tsukuba, Ibaraki 305-0801, Japan\\
$^6$MoE Frontiers Science Center for Rare Isotopes, Lanzhou University, Lanzhou 730000, China}

\begin{abstract}

We investigate the resonance phenomenon in deeply bound charmoniumlike  states. By interpreting the $Y(4220)$, $Y(4360)$, and $\psi(4415)$ states as deeply $S$-wave bound $D \bar D_1$, $D^* \bar D_1$, and $D^* \bar D_2^*$ molecules, respectively, we robustly predict the existence of $4$ $S$-wave deeply bound $C$-parity partner states and $13$ $P$-wave charmoniumlike  molecular resonances. These states, formed by exchanging the light mesons and pinpointed through the complex scaling method, include $3$ $S$-wave bound states with exotic quantum numbers  $I(J^{PC})$= $0(1^{-+})$ and $5$ $P$-wave resonances with exotic $0(0^{-+})$ and $0(2^{+-})$.  We expect future experiments to search for such exotic molecular resonances, which will not only enrich the ongoing construction of ``Particle Zoo 2.0" but also help to scrutinize the hadronic molecular assignments of the $Y$-family particles including the $Y(4220)$, $Y(4360)$, and $\psi(4415)$ states. 
\end{abstract}

\maketitle

\noindent{\it Introduction.---}The observation of the $X(3872)$ in 2003 marked a milestone in the exploration of hadron spectroscopy. Since then, numerous exotic hadronic states beyond conventional conventional $q\bar q$ mesons and $qqq$ baryons have been discovered with advancements in experimental precision, including the charmoniumlike  $XYZ$ states and the $P_c$ pentaquarks. Despite extensive theoretical analyses,  their inner structures remain mysterious due to the lack of experimental data.  (see review articles \cite{Liu:2013waa,Hosaka:2016pey,Chen:2016qju,Richard:2016eis,Lebed:2016hpi,Olsen:2017bmm,Guo:2017jvc,Brambilla:2019esw,Liu:2019zoy,Chen:2022asf,Meng:2022ozq,Amsler:2004ps,Swanson:2006st,Godfrey:2008nc,Yamaguchi:2019vea,Albuquerque:2018jkn,Yuan:2018inv,Ali:2017jda,Dong:2017gaw,Faccini:2012pj,Drenska:2010kg,Pakhlova:2010zza}). 
The $c \bar c q\bar q$  states are particularly challenging to understand. One significant challenge is that the mass spectra of  $c\bar c q\bar q$ states importantly overlap with those of the excited ${c}\bar{c}$ configurations. Moreover, many of them  share the identical  $J^{PC}$ quantum numbers. The $q \bar q$ component can be easily excited within the excited $c \bar c$ system. These complicate distinguishing among various theoretical interpretations of these exotic states such as conventional excited charmonium, hadronic molecules, hybrids, or other exotic configurations. For instance, the $X(3872)$, $X(4140)$, and $Y$-family states (e.g., $Y(4220)$, $Y(4360)$) with $J^{PC}=1^{--}$,  require distinguishing features to differentiate these theoretical scenarios. It is the dilemma of identifying exotic hadron from $XYZ$ data.

To disrupt the status quo, we can draw on charmoniumlike  states $Y(4220)$, $Y(4360)$, and $\psi(4415)$. Previously, Close and Downum found that the $D^* \bar D_1(2430)$ bound state\footnote{In the {\textit{Review of Particle Physics}} (RPP) \cite{Workman:2022ynf}, $D_1(2430)$ and $D_1(2420)$ are  identified as wide and narrow charmed mesons, respectively. $D_1(2430)$ is difficult to form molecules with  $D^{(*)}$  due to its substantial decay width. Our study focuses on molecules composed of $D^{(*)}$ and $D_1(2420)/D_2^*(2460)$, using the shorthand $D^{(*)}\bar{D}_1$+c.c./$D^{*}\bar{D}_2^*$+c.c. for these systems. Here, $D_1$ and $D_2^*$ denote $D_1(2420)$ and $D_2^*(2460)$, respectively \cite{Workman:2022ynf}.}, with the binding energy around $\mathcal{O}(100)$ MeV \cite{Close:2009ag,Close:2010wq}, can be assigned to the observed charmoniumlike  state $Y(4260)$ or $Y(4360)$ \cite{Close:2009ag,Close:2010wq}. In Ref. \cite{Ji:2022blw}, the authors proposed that the $Y(4220)$, $Y(4360)$, and $\psi(4415)$ states are deeply bound molecules of  $D\bar{D}_1$, $D^{*}\bar{D}_1$, and $D^{*}\bar{D}_2^{*}$  with binding energies in the range of $\mathcal{O}(50)$ to $\mathcal{O}(100)$ MeV, respectively. Additionally, they predicted a $D^{*}\bar{D}_1$ molecular state with the exotic quantum number of $J^{PC}=0^{--}$ \cite{Ji:2022blw}.

In this Letter, we propose a novel approach to understanding these states by predicting $P$-wave resonances from deeply bound molecular states. Notably, considering a $P$-wave excitation in the two constituent mesons introduces larger number of the exotic $J^{PC}$ quantum numbers forbidden in conventional $Q\bar Q$ systems, helping to avoid the coupled-channel effects  between $Q\bar Q$ and $Q\bar Q q\bar q$.  We demonstrate that if the renowned exotic states $Y(4220)$, $Y(4360)$, and $\psi(4415)$ are the deeply bound charmoniumlike  molecular states  as suggested in Refs. \cite{Close:2009ag,Close:2010wq,Ji:2022blw}, there will exist correspondingly the molecular counterparts with exotic $J^{PC}=0(1^{-+})$ ($S$-wave bound state),  ${0^{+-}}$ and $2^{+-}$ ($P$-wave resonances). The existence of resonances in the  $D\bar{D}_1$, $D^{*}\bar{D}_1$, and $D^{*}\bar{D}_2^{*}$ systems necessitates predicting their characteristic spectra.
Interpreting  the $Y$ and $\psi$ states as the molecules  allowed us to establish reliable meson-meson interactions through light meson exchange.  Utilizing the complex scaling  method (CSM) to identify poles, we systematically predict the mass spectrum of deeply bound $S$-wave charge-conjugation states alongside $P$-wave resonances as counterparts of the $Y(4220)$, $Y(4360)$, $\psi(4415)$, and $\psi_0(4360)$ for the first time.  These states provide novel insights into  the scenario of the deeply bound charmoniumlike  molecules proposed in previous studies (see Refs. \cite{Close:2009ag,Close:2010wq,Ji:2022blw}), and  paves the way for new spectroscopic research in hadron physics.

This methodology  can be easily  extended to study other exotic states  and enrich our understanding of resonance phenomena in hadronic physics. The resonance phenomena also play a crucial role  in nuclear physics, where the discoveries of the numerous exotic nuclei, such as $^{11}$Li \cite{Tanihata:1985psr}, $^{31}$Ne \cite{Nakamura:2009zzh}, and $^{22}$C \cite{Tanaka:2010zza}, have highlighted the pivotal role of resonances in the formation of the halo, giant halo, and deformed halo structures in the  exotic nuclei \cite{Meng:1996zz,Poschl:1997ky,Sandulescu:2003jw,Meng:1996tt,Meng:2002ps,Terasaki:2006tw,Grasso:2006es,Hamamoto:2009wj,Zhou:2009sp}. Identifying and analyzing resonances  not  only advances hadronic physics but also underscores the importance of resonance phenomena across different areas of physics.

\noindent{\it Framework.---}For the  $D \bar D_1$, $D^* \bar D_1$, and $D^* \bar D_2^*$ systems, labeled as $\mathcal{A}\overline{\mathcal{B}}$ \footnote{Here and after, the notations $\mathcal{A}$ and $\mathcal{B}$ denote  different charmed mesons. }, the flavor wave functions with the isospin $|I, I_3\rangle$ are given by
\begin{align} \label{eq:wf}
    &|0,0\rangle=\dfrac{1}{2}\left[\left(\mathcal{A}^{0}\overline{\mathcal{B}}^{0}+\mathcal{A}^{+}{\mathcal{B}}^{-}\right)+c\left(\mathcal{B}^{0}\overline{\mathcal{A}}^{0}+\mathcal{B^+}{\mathcal{A}}^{-}\right)\right],\nonumber\\
    &|1, 1\rangle=\dfrac{1}{\sqrt{2}}\left(\mathcal{A}^{+}\overline{\mathcal{B}}^{0}+c\mathcal{B}^{+}\overline{\mathcal{A}}^{0}\right),\nonumber\\
    &|1,0\rangle=\dfrac{1}{2}\left[\left(\mathcal{A}^{0}\overline{\mathcal{B}}^{0}-\mathcal{A}^{+}{\mathcal{B}}^{-}\right)+c\left(\mathcal{B}^{0}\overline{\mathcal{A}}^{0}-\mathcal{B^+}{\mathcal{A}}^{-}\right)\right],\nonumber\\
    &|1, -1\rangle=\dfrac{1}{\sqrt{2}}\left({\mathcal{A}}^{0}\mathcal{B}^{-}+c{\mathcal{B}}^{0}\mathcal{A}^{-}\right),
\end{align}
where  the $C$ parity of $\mathcal{A}\overline{\mathcal{B}}$ system is defined as $C=cx_1x_2(-1)^{L+S+s_1+s_2}$  with the factor $c=\pm1$ \cite{Liu:2007bf,Liu:2008xz,Liu:2008fh,Liu:2008tn,Sun:2012sy,Wang:2020dya,Li:2015exa,Li:2013bca,Hu:2010fg,Shen:2010ky,Dong:2021juy,Liu:2013rxa,Chen:2015add}. $L$ and $S$ denote the orbital angular momentum and spin of the system, respectively. $s_1$ and $s_2$ represent the spin, and $x_1=\pm1$ and $x_2=\pm1$ indicate the charge-conjugation conventions of charmed mesons $\mathcal{A}$ and $\mathcal{B}$, respectively. In this work, we adopt the charge-conjugation conventions for the $D$, $D^*$, $D_1$, and $D_2^*$ mesons as specified in Ref. \cite{Ding:2008gr}.

With Eq. \eqref{eq:wf}, the effective potential $\mathcal{V}_E(\bm{q})$ can be calculated as 
\begin{eqnarray}
\mathcal{V}_E(\bm{q})=\mathcal{V}_E^{D}(\bm{q})+c\,\mathcal{V}_E^{C}(\bm{q}).
\end{eqnarray}
Here, $\mathcal{V}_E^{D}(\bm{q})$ denotes the effective potentials from the direct progress $\mathcal{A}(\bm{p_1})\overline{\mathcal{B}} (\bm{p_2})\rightarrow \mathcal{A}(\bm{p_3})\overline{\mathcal{B}} (\bm{p_4})$, and $\mathcal{V}_E^{C}(\bm{q})$ from the crossed one $\mathcal{A}(\bm{p_1})\overline{\mathcal{B}} (\bm{p_2})\rightarrow \overline {\mathcal{A}}(\bm{p_3}){\mathcal{B}} (\bm{p_4})$, respectively.

The effective potentials between $D \bar D_1$, $D^* \bar D_1$, and $D^* \bar D_2^*$ systems are established by exchange of the $\pi$, $\sigma$/$\eta$, and $\rho$/$\omega$. The relevant effective Lagrangians can be constructed under the heavy-quark symmetry, the chiral symmetry, and the hidden local symmetry \cite{Burdman:1992gh,Wise:1992hn,Yan:1992gz,Ding:2008gr}
\begin{eqnarray}
    {\mathcal L}_{{H}-{H}-{L}}&=&g_{\sigma}\left\langle H^{(Q)}_a\sigma\overline{H}^{(Q)}_a\right\rangle+g_{\sigma}\left\langle \overline{H}^{(\overline{Q})}_a\sigma H^{(\overline{Q})}_a\right\rangle\nonumber\\
    &&+ig\left\langle H^{(Q)}_b{\mathcal A}\!\!\!\slash_{ba}\gamma_5\overline{H}^{\,({Q})}_a\right\rangle+ig\left\langle \overline{H}^{(\overline{Q})}_a{\mathcal A}\!\!\!\slash_{ab}\gamma_5 H^{\,(\overline{Q})}_b\right\rangle\nonumber\\
    &&+\left\langle iH^{(Q)}_b\left(\beta v^{\mu}({\mathcal V}_{\mu}-\rho_{\mu})+\lambda \sigma^{\mu\nu}F_{\mu\nu}\right)_{ba}\overline{H}^{\,(Q)}_a\right\rangle,\nonumber\\
    {\mathcal L}_{{T}-{T}-{L}}&=&g^{\prime\prime}_{\sigma}\left\langle T^{(Q)\mu}_a\sigma\overline{T}^{(Q)}_{a\mu}\right\rangle+g^{\prime\prime}_{\sigma}\left\langle\overline{T}^{(\overline{Q})\mu}_a\sigma T^{(\overline{Q})}_{a\mu}\right\rangle\nonumber\\
    &&+ik\left\langle T^{\,(Q)\mu}_b{\mathcal A}\!\!\!\slash_{ba}\gamma_5\overline{T}^{(Q)}_{a\mu}\right\rangle+ik\left\langle\overline{T}^{\,(\overline{Q})\mu}_a{\mathcal A}\!\!\!\slash_{ab}\gamma_5T^{(\overline{Q})}_{b\mu}\right\rangle\nonumber\\
    &&+\left\langle iT^{\,(Q)}_{b\lambda}\left(\beta^{\prime\prime} v^{\mu}({\mathcal V}_{\mu}-\rho_{\mu})+\lambda^{\prime\prime}\sigma^{\mu\nu}F_{\mu\nu}\right)_{ba}\overline{T}^{(Q)\lambda}_{a}\right\rangle\nonumber\\
    &&-\left\langle i\overline{T}^{\,(\overline{Q})}_{a\lambda}\left(\beta^{\prime\prime} v^{\mu}({\mathcal V}_{\mu}-\rho_{\mu})-\lambda^{\prime\prime}\sigma^{\mu\nu}F_{\mu\nu}\right)_{ab}T^{(\overline{Q})\lambda}_{b}\right\rangle,\nonumber\\
    {\mathcal L}_{{H}-{T}-{L}}&=&\frac{h^{\prime}_{\sigma}}{f_{\pi}}\left[\left\langle T^{(Q)\mu}_a\partial_{\mu}\sigma\overline{H}^{(Q)}_b\right\rangle+\left\langle\overline{H}^{(\overline{Q})}_a\partial_{\mu}\sigma T^{(\overline{Q})\mu}_b\right\rangle\right]\nonumber\\
    &&+\left[i\left\langle T^{(Q)\mu}_b\left(\frac{h_1}{\Lambda_{\chi}}D_{\mu}{\mathcal A}\!\!\!\slash+\frac{h_2}{\Lambda_{\chi}}D\!\!\!\!/ {\mathcal A}_{\mu}\right)_{ba}\gamma_5\overline{H}^{\,(Q)}_a\right\rangle\right]\nonumber\\
    &&+\left[i\left\langle\overline{H}^{\,(\overline{Q})}_a\left(\frac{h_1}{\Lambda_{\chi}}{\mathcal A}\!\!\!\slash\stackrel{\leftarrow}{D_{\mu}'}+\frac{h_2}{\Lambda_{\chi}}{\mathcal A}_{\mu}\stackrel{\leftarrow}{D\!\!\!\slash'}\right)_{ab}\gamma_5T^{(\overline{Q})\mu}_b\right\rangle\right]\nonumber\\
    &&+\left[\left\langle T^{(Q)\mu}_b\left(i\zeta_1({\mathcal V}_{\mu}-\rho_{\mu})+\mu_{1}\gamma^{\nu}F_{\mu\nu}\right)_{ba}\overline{H}^{\,(Q)}_a\right\rangle\right]\nonumber\\
    &&-\left[\left\langle\overline{H}^{\,(\overline{Q})}_a\left(i\zeta_1({\mathcal V}_{\mu}-\rho_{\mu})-\mu_1\gamma^{\nu}F_{\mu\nu}\right)_{ab}T^{(\overline{Q})\mu}_b\right\rangle\right]\nonumber\\[2mm]
    &&+h.c.,
\end{eqnarray}
where the superfields $H^{(Q)}_a$ and $T^{(Q)\mu}_a$ represent the $S$-wave $(D,D^*)$ and $P$-wave $(D_1,D_2^*)$ doublets in the heavy-quark limit, respectively. The key elements, including axial current $\mathcal{A}_\mu$, vector current ${\cal V}_{\mu}$, the vector meson field $\rho_{\mu}$, and the vector meson strength tensor $F_{\mu\nu}$, are specified in Ref. \cite{Ding:2008gr}. The  coupling constants can be derived from both experimental data and theoretical models, summarized in Supplemental Material \cite{{wang:sup}}. 

With the Breit approximation \cite{Breitapproximation}, we acquire the effective potentials $\mathcal{V}(\bm{q})$ using the scattering amplitude $\mathcal{M}(\bm{q})$.
\begin{eqnarray}
    \mathcal{V}^{h_1h_2\to h_3h_4}_E(\bm{q})=-\frac{\mathcal{M}^{h_1h_2\to h_3h_4}(\bm{q})}{4\sqrt{\smash[b]{m_{h_1}m_{h_2}m_{h_3}m_{h_4}}}}\mathcal{F}^2(q^2,m_E^2),
\end{eqnarray}
where $m_{h_i}$ are the interacting hadron masses. A monopole form factor is introduced to compensate the off-shell effect of the exchanged mesons and describe the structure effect of each effective vertex $\mathcal{F}(q^2,m_E^2) = \frac{\Lambda^2-m_E^2}{\Lambda^2-q^2}$ \cite{Tornqvist:1993ng,Tornqvist:1993vu,Wang:2020dya}.
Here, $\Lambda$ is a cutoff parameter, $m_E$ and $q$ are the mass and four momentum of the exchanged light mesons, respectively. These formulations allow comprehensive expressions of the $D \bar D_1$, $D^* \bar D_1$, and $D^* \bar D_2^*$ effective potentials as detailed in the Supplemental Material \cite{{wang:sup}}.

To explore bound and resonant states, we apply the CSM \cite{Myo:2014ypa,Myo:2020rni,Moiseyev:1998gjp},  which has been utilized in several recent studies to investigate potential hadronic molecules \cite{Wang:2023ovj,Ren:2023pip,Chen:2023eri,Cheng:2022qcm,Cheng:2022vgy,Cheng:2023vyv}. The CSM introduces a dilation transformation operator $U(\theta)$ affecting the coordinate $\bm r$ and its conjugate momentum $\bm p$,
\begin{eqnarray}\label{transformation1}
    U(\theta)\bm{r}U^{-1}(\theta)=\bm{r}e^{i\theta}, \quad U(\theta)\bm{p}U^{-1}(\theta)=\bm{p}e^{-i\theta},
\end{eqnarray}
where $\theta$ is the scaling angle. The Schr\"odinger equation is then transformed as
\begin{align}\label{eq:transformation2}
    &\frac{p^2 e^{-2i\theta}}{2\mu}\psi_{L}^{\theta}(p)+\int \frac{e^{-3i\theta}p^{\prime 2}\mathrm{d} p'}{(2\pi)^3}V_{LL}^{JS}(p e^{-i\theta},p'e^{-i\theta})\psi_{L}^{\theta}(p')\nonumber\\
    &=E^{\theta}\psi_{L}^{\theta}(p),
\end{align}
where $E^{\theta}$ is the complex eigenvalue, and $\mu$ is the reduced mass of interacting hadrons. $V_{LL}^{JS}(p,p')$ is the partial wave potential. The numerical calculations  are described in Supplemental Material \cite{{wang:sup}}.

\noindent{\it $S$-wave counterpart states---} Assuming  the $Y(4220)$, $Y(4360)$, and $\psi(4415)$ states are deeply bound $D \bar D_1$, $D^* \bar D_1$, and $D^* \bar D_2^*$ molecules with $I(J^{PC})=0(1^{--})$ , respectively \cite{Close:2009ag,Close:2010wq,Ji:2022blw}, we first determine their cutoffs $\Lambda$ using their binding energies. We then examine  their $C$-parity counterparts with $I(J^{PC})=0(1^{-+})$, summarizing the cutoffs and results in Table \ref{cutoff}.

\begin{table}[!htbp]
    \caption{The cutoffs are determined building on the hypothesis that the $Y(4220)$, $Y(4360)$, $\psi(4415)$, and $\psi_0(4360)$ are $C=-$ deeply bound molecules with the binding energies $E^-$ \cite{Ji:2022blw}. Additionally, the binding energies $E^+$ of the $C=+$ counterparts  are given with the same cutoff values.}\label{cutoff}
    \renewcommand\tabcolsep{0.15cm}
    \renewcommand{\arraystretch}{1.50}
    \centering
    \begin{tabular}{c|ccccc}
        \toprule[1.0pt]
        \toprule[1.0pt]
        Systems                 &$I(J^P)$   &Molecules $[E^-\ (\rm{MeV})]$    &$E^+$ (MeV) &$\Lambda$ (\rm{GeV})    \\
        \midrule[0.75pt]
        $D^{*}\bar{D}^{*}_{2}$  &$0(1^-)$   &$\psi(4415)\ [-49]$            &$-50$      &$1.11$                  \\
        $D^{*}\bar{D}_{1}$      &$0(1^-)$   &$Y(4360)\ [-62]$               &$-60$      &$1.29$                  \\
        $D\bar{D}_{1}$          &$0(1^-)$   &$Y(4220)\ [-67]$               &$-100$     &$2.23$                  \\
        $D^{*}\bar{D}_{1}$      &$0(0^-)$   &$\psi_0(4360)\ [-63]$          &$-60$      &$1.11$                  \\
        \bottomrule[1.00pt]
        \bottomrule[1.00pt]
    \end{tabular}
\end{table}

Drawing from earlier one-boson-exchange model (OBE) studies on deuteron, the cutoff value for potential hadronic molecules is expected to be typically around $1$ GeV  \cite{Machleidt:1987hj,Epelbaum:2008ga,Esposito:2014rxa,Chen:2016qju}. In Table \ref{cutoff}, the expectation is particularly notable for the $Y(4360)$, $\psi(4415)$, and $\psi_{0}(4360)$ states. However,  a larger cutoff value is needed to reproduce the mass of the $Y(4220)$ state as the $D \bar D_1$ bound state with $I(J^{PC})=0(1^{--})$.  This study mainly focuses on the implications of the existence of these deeply bound states. Therefore, we do not try to decode the inner dynamics and use the larger cutoff values to explore the existence of resonances.

 Our results show that the $C=+$ states exhibit strong attractive potentials, leading to deeply bound $I(J^{PC})=0(1^{-+})$ and $I(J^{PC})=0(0^{-+})$ molecules, as shown in Table \ref{cutoff}. Their binding energies closely match those of the  $Y$ and $\psi$ states, except for the $I(J^{PC})=0(1^{-+})$ $D \bar D_1$ states with a large binding energy up to $100$ MeV. Notably, the interaction differences between  $C=+$ and $C=-$ counterparts stems from the interference of $\mathcal{V}_E^{D}(\bm{q})$ and  $\mathcal{V}_E^{C}(\bm{q})$. The similar binding energies in the $D^*\bar D_{1}$ and $D^*\bar D_{2}$ systems suggest the dominance of direct-diagram potentials over cross-diagram ones.  
 Meanwhile, we do not find any $S$-wave radially excited resonances corresponding to the $S$-wave bound systems.

For the $S$-wave $I=1$ systems, we found no bound states, which is consistent with the interaction differences between  $I=0$ (isoscalar) and $I=1$ (isovector) systems. This difference primarily arises from the $\rho$ and $\pi$ meson exchanges, guided by the isospin factor $I(I+1)-\frac{3}{2}$, making $I=0$ interactions more attractive negative triple of $I=1$  systems. Additionally, the $\rho$ exchange effects are almost canceled out by the opposing $\omega$ meson exchange impacts. This indicates that the $I=0$ molecules  form more readily than  their $I=1$ counterparts.

\noindent{\it $P$-wave resonances---}Turning to $P$-wave excitations, the important centrifugal barrier may drive the $S$-wave deeply bound states into resonances \cite{herzberg:1950mol}. With the CSM, we find the existence of the resonance solutions for the  $P$-wave $I=0$ $D\bar{D}_{1}$, $D^{*}\bar{D}_{1}$, and $D^{*}\bar{D}^{*}_{2}$ systems when the corresponding  $S$-wave $I=0$ systems form deeply bound states  $Y(4220)$, $Y(4360)$, $\psi(4415)$,  and $\psi_{0}(4360)$, respectively. Conversely, for the $S$-wave $I=1$ systems, no bound states are observed,  and the corresponding $P$-wave $I=1$ resonance candidates also do not exist. This indicates that the presence of the $S$-wave deeply bound states hints at the possible $P$-wave resonances, and the orbital angular momentum plays a crucial role in resonance formation, as does predissociation by rotation of the diatomic molecules \cite{herzberg:1950mol}.

\begin{figure}[htbp]
    \centering
    \includegraphics[width=0.48\textwidth]{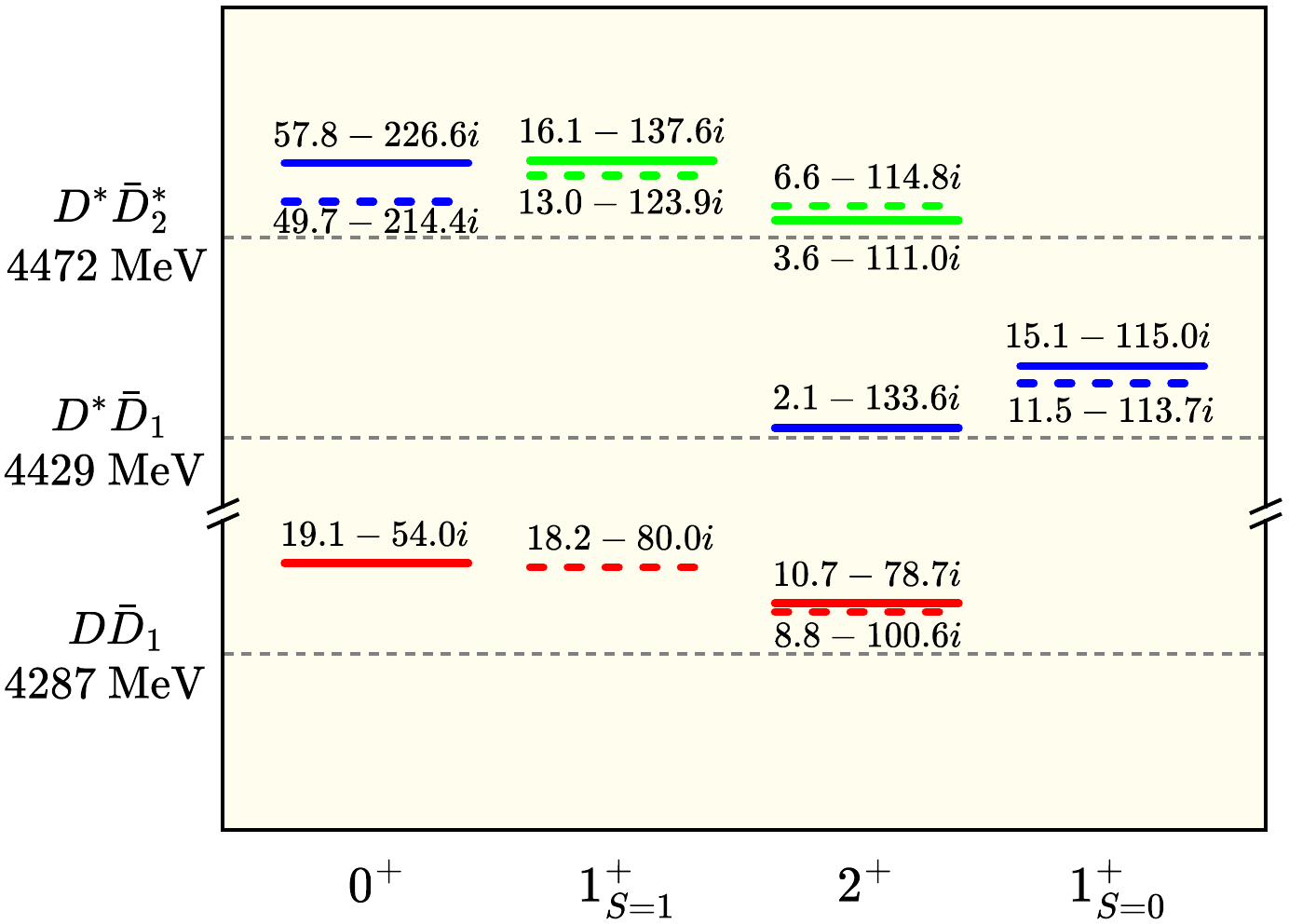}
    \caption{The characteristic spectrum ($E_r-i\Gamma_r/2$) of the $P$-wave $I=0$ molecular resonances for the $D\bar{D}_{1}$ (red), $D^{*}\bar{D}_{1}$ (blue), and $D^{*}\bar{D}^{*}_{2}$ (green), based on deeply bound molecular hypothesis. Here, the $C=-$ and $C=+$ resonances are denoted by solid and dashed lines, respectively. }\label{spectrum}
\end{figure}

Experimentally, the $Z_c(4430)^+$ with $I(J^P)=1(1^+)$ has been observed in the $\psi(2S) \pi^+$ invariant mass distribution \cite{LHCb:2014zfx,Belle:2007hrb,Belle:2013shl}.
It was  a potential candidate for the $P$-wave $I=1$ $D^*\bar D_1(2430)$ \cite{Barnes:2014csa} and $D^*\bar D_1$ \cite{He:2017mbh} molecules. Unlike Ref.~\cite{He:2017mbh}, which focused on the cross diagram of the $\pi$ exchange, our study also incorporates the more significant direct $\pi$ exchange alongside the vector and scalar exchanges. The absence of the $P$-wave $I=1$ molecules in our study suggests that identifying $Z_c(4430)^+$ as the $P$-wave $D^*\bar D_1$ molecules is not viable when considering the $Y(4360)$ as the $S$-wave $I(J^{PC})=0\left(1^{--}\right)$ $D^*\bar D_1$ molecule.

In Fig. \ref{spectrum}, we present the characteristic spectrum of the $P$-wave $I=0$ $D\bar{D}_{1}$, $D^{*}\bar{D}_{1}$, and $D^{*}\bar{D}^{*}_{2}$ molecular resonances\footnote{The $D^*\bar D_1$ systems with total spin $S=0$ and $S=1$ can form identical quantum number $J^{PC}=1^{+\pm}$. To avoid confusion, we will use the notation $J^{PC}_{S=0}$ /$J^{PC}_{S=1}$ to label the relevant quantum numbers for the $D^*\bar D_1$ systems. }.
For $C=-$ systems, there exist seven charmoniumlike  molecular resonance candidates: the $D\bar{D}_{1}$ with $J^{PC}=0^{+-}$ and $2^{+-}$, the $D^{*}\bar{D}_{1}$  with $J^{PC}=0^{+-}$, $2^{+-}$ and $1^{+-}_{S=0}$, and the $D^{*}\bar{D}_{2}^{*}$  with $J^{PC}=1^{+-}$ and $2^{+-}$. For $C=+$  systems, the $D\bar{D}_{1}$  with $J^{PC}=1^{++}$ and $2^{++}$, the $D^{*}\bar{D}_{1}$  with $J^{PC}=0^{++}$ and $1^{++}_{S=0}$, and the $D^{*}\bar{D}_{2}^{*}$  with $J^{PC}=1^{++}$ and $2^{++}$ can be considered as charmoniumlike  molecular resonance candidates.

\begin{figure}[htbp]
    \centering
    \includegraphics[width=0.43\textwidth]{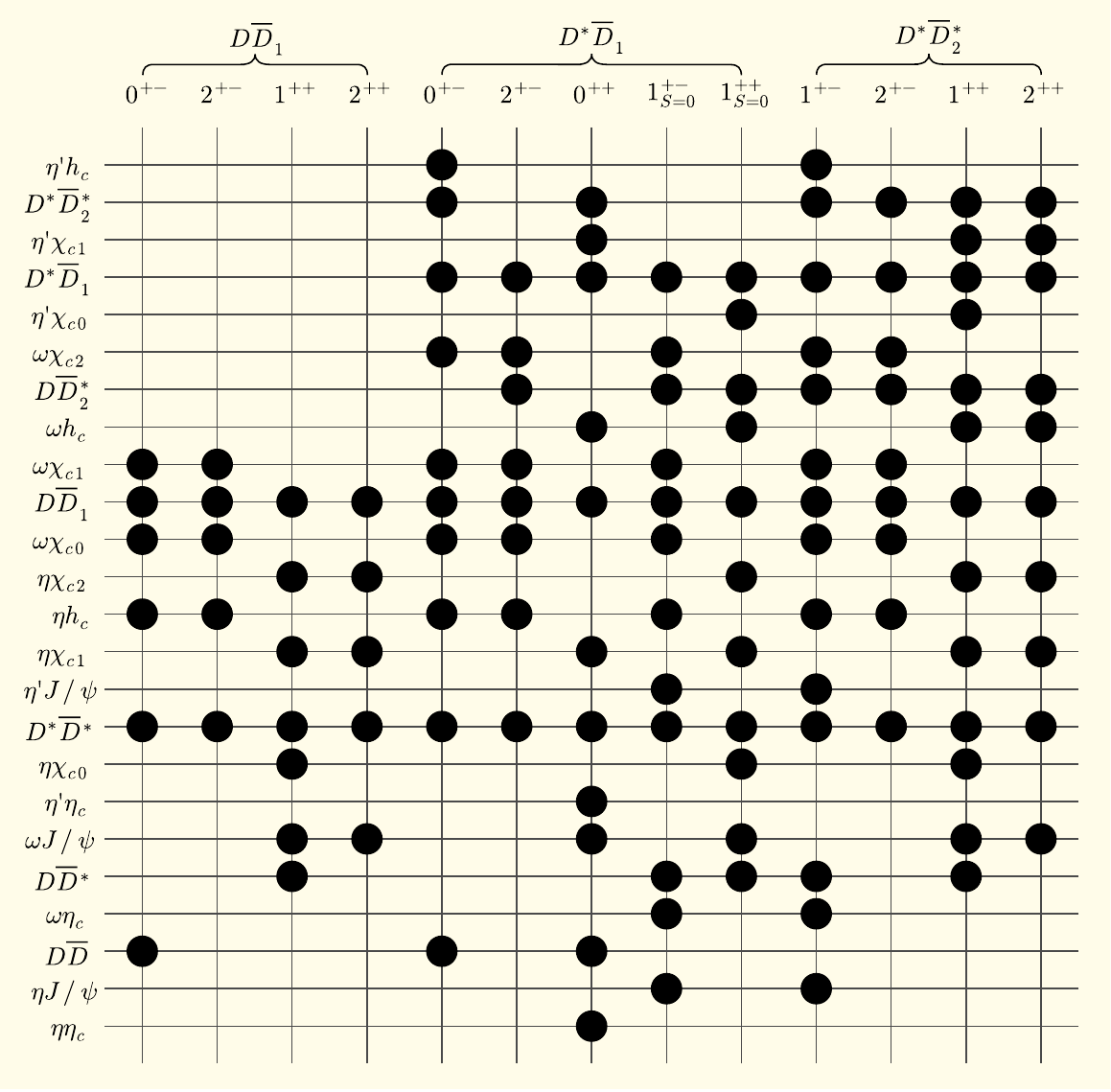}
    \caption{The potential two-body hidden-charm and open-charm decay channels for the obtained $P$-wave isoscalar $D\bar{D}_1$, $D^{*}\bar{D}_1$, and $D^{*}\bar{D}_2^{*}$ charmoniumlike  molecular resonances.}\label{decay}
\end{figure}

As shown in Fig. \ref{spectrum}, the energies of these obtained resonances are close to the corresponding open-charm thresholds, except for the $D^{*}\bar{D}_{1}$ resonances with $I(J^{PC})=0(0^{+\pm})$. Furthermore, their decay widths exceeds $100$ MeV and satisfy the order $\Gamma[D\bar{D}_{1}]<\Gamma[D^{*}\bar{D}_{2}^{*}]<\Gamma[D^{*}\bar{D}_{1}]$ for the identical quantum numbers. The large decay width arise from decays into the scattering states composed of  two constituent heavy mesons $D^{(*)}\bar{D}_{1,2}$.
These broad resonances can also decay into two-body hidden-charm and other open-charm final states, which are shown in Fig. \ref{decay}.  Due to the molecular nature of these resonant states, we expect a larger decay branching ratio for open-charm channels than for hidden-charm channels.  

The substantial decay width of the $P$-wave resonance may cover the region of $D^{(*)}\bar{D}_{1,2}$ channels and suggest the potential coupled-channel effects. However, we did not explicitly include such effects  because our approach mirrors previous investigations of $S$-wave potentials, which also did not consider these effects explicitly. Our results for the $S$-wave interactions were based on experimental data, where the interaction strengths for the $D^{(*)}D_{1,2}$ channel have been  well constrained to describe the $S$-wave deeply bound molecular states. We believe that the essential contributions of coupled-channel effects have been effectively absorbed into the short-range interactions, which are parametrized by the coupling constants and cutoff parameters in our model. Since the $S$-wave and $P$-wave interactions stem from the same underlying potential via the partial wave expansion, the fundamental conclusions about the nature of the $P$-wave states are expected to remain robust, given the accurate description of the $S$-wave mass spectrum.

The conventional charmonia may couple with the predicted resonances sharing identical quantum numbers and comparable masses. However, several resonances have the exotic quantum numbers that distinguish them from conventional charmonia, such as the  $D\bar{D}_{1}$ and  $D^{*}\bar{D}_{1}$ resonances with $I(J^{PC})=0(0^{+-})$ and $0(2^{+-})$, and the $D^{*}\bar{D}_{2}^{*}$ resonance with $I(J^{PC})=0(2^{+-})$. This distinction ensures their existence and mass spectra remain robust against coupled-channel influences. Among these, the $D\bar{D}_{1}$ resonance with $I(J^{PC})=0(0^{+-})$ is notable for the narrowest decay width of approximately $100$ MeV, making it a particularly promising candidate for experimental observation.

\begin{figure}[htbp]
    \centering
    \includegraphics[width=0.45\textwidth]{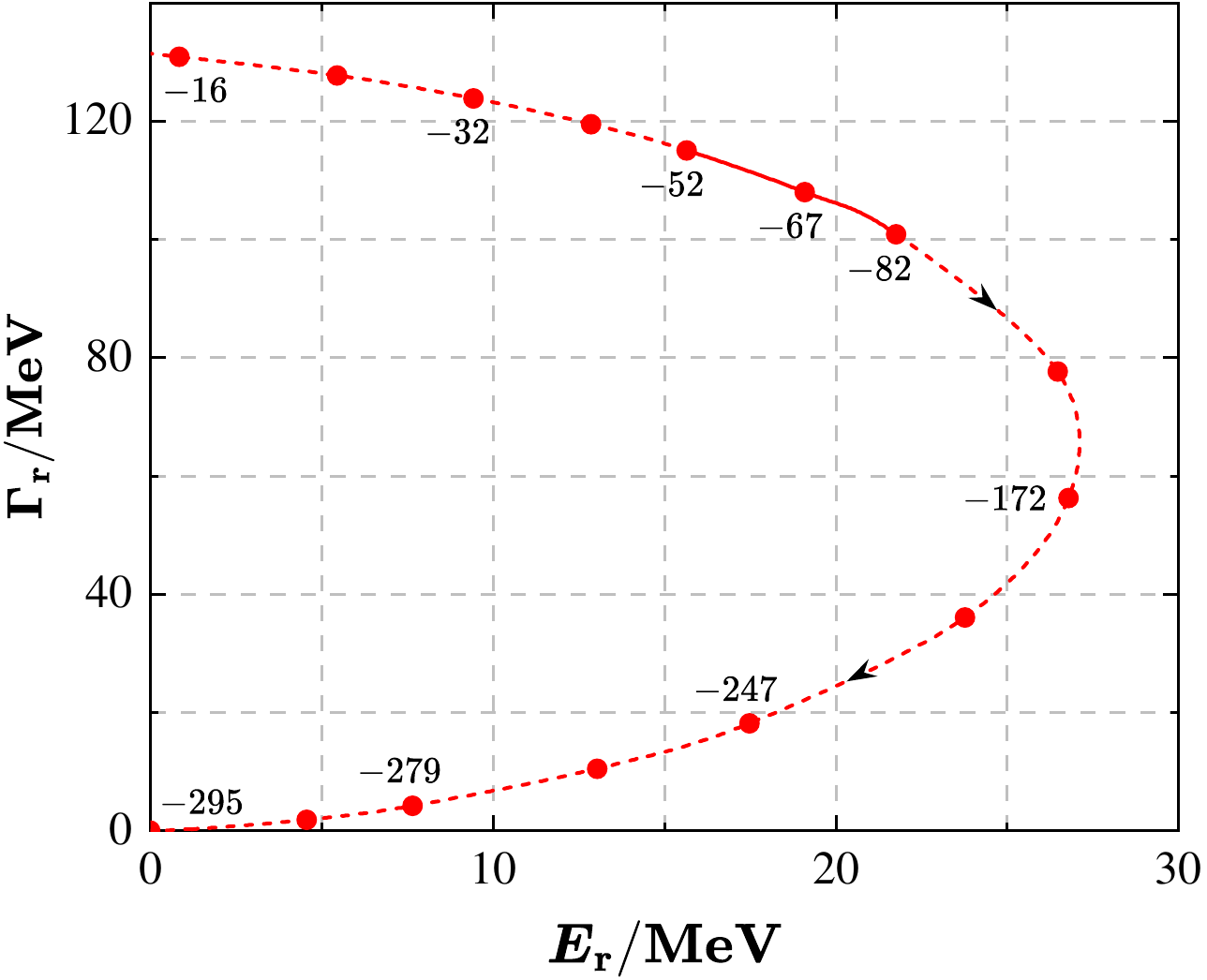}
    \caption{The trajectories for the $P$-wave $D\bar D_1$ resonance with $I(J^{PC})=0(0^{+-})$ vary with the binding energy of the corresponding $S$-wave molecule with $I(J^{PC})=0(1^{--})$. Here, the solid line indicates the $Y(4220)$ as the $D\bar D_1$ molecule considering the experimental uncertainty.}\label{r_para_b}
\end{figure}

To demonstrate the correlation between the existence of the resonances and the corresponding deeply bound molecules, we display how the $P$-wave $D\bar D_1$ resonance with $I(J^{PC})=0(0^{+-})$ varies with the binding energy of the $S$-wave $D\bar D_1$ molecule with $I(J^{PC})=0(1^{--})$ in Fig. \ref{r_para_b}. When the binding energy of the $S$-wave $D\bar D_1$ molecule  exceeds $15$ MeV, the $P$-wave $D\bar D_1$ resonance appears.  As the binding energy further increases, the resonance width will decrease. The real part $E_r$ first increases then reaches a maximum, decreases, and finally vanishes, where the resonance transitions to a bound state. Furthermore, if the $Y(4220)$ is identified as the $S$-wave deeply bound $D\bar D_1$ molecule, the corresponding $P$-wave $D\bar D_1$ resonance with $I(J^{PC})=0(0^{+-})$ must exist.

{\noindent{\it Summary.---}}Our calculations, focused on the deeply bound molecular states, uncovers the $P$-wave resonance phenomena in molecules with substantial binding energies. Such phenomena might be more general across exotic hadronic states. Specifically, our numerical results show that deeply bound molecular assignments of the $Y(4220)$, $Y(4360)$, $\psi(4415)$ and a theoretically predicted $\psi_0(4360)$ naturally lead to the emergence of four deeply bound $S$-wave $I=0$ molecules $D \bar D_1$, $D^* \bar D_1$, $D^* \bar D_2^*$ with positive $C$ parities, including one with  $I(J^{PC})=0(0^{-+})$ and three with exotic $I(J^{PC})=0(1^{-+})$. We do not find the existence of $S$-wave excited resonances. Furthermore, we predict $P$-wave $I=0$ resonances:  $D\bar{D}_{1}$ with $J^{PC}=0^{+-}$, $1^{++}$, and $2^{+\pm}$, the $D^{*}\bar{D}_{1}$  with $J^{PC}=0^{+\pm}$, $2^{+-}$, and $1^{+\pm}_{S=0}$, and the  $D^{*}\bar{D}_{2}^{*}$  with $J^{PC}=1^{+\pm}$ and $2^{+\pm}$. Despite potential shifts in mass spectra due to coupled-channel effects with excited charmonium, predictions for $S$-wave states with  exotic $I(J^{PC})=0(1^{-+})$ and $P$-wave resonances with  exotic $I(J^{PC})=0(0^{+-})$, $0(2^{+-})$  are robust. These $P$-wave resonances, characterized by large decay widths, suggest two-body hidden-charm and open-charm decay channels as promising avenues for experimental exploration. 

Additionally, our findings rule out the existence of  $S$-wave $I=1$ bound states or radially excited resonances and the corresponding  $P$-wave $I=1$ resonances. This challenges the
interpretation of the $I=1$ $Z_c(4430)^+$  as the 
$P$-wave $D^*\bar D_1$ molecular structures.

The predicted resonances, if confirmed, would not only enrich the spectrum of charmoniumlike  molecular resonances, but also help to examine the hadronic molecular nature of the $Y(4220)$, $Y(4360)$, and $\psi(4415)$ states \cite{Close:2009ag,Close:2010wq,Ji:2022blw}.
We therefore strongly expect future experiments to search for these predicted charmoniumlike  molecular resonances near the $D\bar{D}_{1}$, $D^{*}\bar{D}_{1}$, and $D^{*}\bar{D}^{*}_{2}$ thresholds.

\noindent{\it Acknowledgement.---} This work is supported by the National Natural Science Foundation of China under Grant No. 12335001, No. 12247101, and No. 12247155, the National Key Research and Development Program of China under Contract No. 2020YFA0406400, and the 111 Project under Grant No. B20063, the Fundamental Research Funds for the Central Universities, and the project for top-notch innovative talents of Gansu province. F.-L.W. is also supported by the China Postdoctoral Science Foundation under Grant No. 2022M721440. G.-J.W is supported  by
the KAKENHI under Grants No.~23K03427 and 	No. 24K17055.

\end{document}